\documentclass[12pt]{article}
\usepackage{times}
\usepackage{graphics}
\newenvironment{sciabstract}{%
\begin{quote} \bf}
{\end{quote}}

\newcounter{lastnote}

\newcommand{\dg}{^\circ}

\title{Discovery of Pulsed OH Maser Emission Stimulated by a Pulsar}

\author{Joel M. Weisberg,$^{1,2,3\ast}$ Simon Johnston,$^{2,1}$ B\"arbel 
Koribalski,$^{1}$ Snezana Stanimirovi{\' c}$^{4}$ \\
\\
\normalsize{$^{1}$Australia Telescope National Facility / CSIRO, P. O. Box 76, 
Epping, NSW 1710, Australia}\\
\normalsize{$^{2}$School of Physics, University of Sydney, 
NSW 2006,  Australia}\\
\normalsize{$^{3}$Department of Physics and Astronomy, Carleton College, 
Northfield, MN 55057, USA}\\
\normalsize{$^{4}$Department of Astronomy, University of California, 
Berkeley, CA 94720, USA}\\
\\
\normalsize{$^\ast$To whom correspondence should be addressed; E-mail:  
jweisber@carleton.edu.}
}

\date{Accepted by Science, May 2005}

\begin{document} 



\maketitle


\begin{sciabstract}
Stimulated emission of radiation has not  been directly observed in 
astrophysical situations up to this time. Here we demonstrate that  
photons from pulsar B1641$-$45 stimulate  pulses of  excess 1720 MHz line 
emission in an interstellar OH  cloud.  As this stimulated emission is driven 
by the pulsar, it varies on a few millisecond timescale, orders of magnitude 
shorter than the quickest OH maser variations previously detected. 

Our  1612 MHz spectra are inverted copies of the 1720 MHz spectra.  This
``conjugate line'' phenomenon enables us to constrain the properties
of the interstellar OH line-producing gas.

We also show that pulsar signals suffer significantly deeper 
OH absorption than do other background sources; confirming earlier tentative 
findings  that OH clouds are clumpier on small scales than 
neutral hydrogen clouds.

\end{sciabstract}

Pulsars have proved to be outstanding tools for study of the interstellar 
medium. Their pulsed signals suffer a variety of modifications as they 
propagate through the
interstellar medium (ISM), revealing extensive
information about the global distribution and physical properties of the 
intervening material.  The  tiny sizes of pulsars ensure that their signals 
probe 
very small transverse  scales in the ISM ({\it{1,2}}).  Another virtue of 
pulsar ISM measurements  is that the 
pulsar cycles rapidly on and off so that observations may be made 
contemporaneously in both the presence and
absence of the pulse, and the properties of the medium can be precisely 
compared in these two states.
For example, comparison of neutral hydrogen (HI) spectra acquired during and 
between pulses leads to  pulsar absorption spectra that can be used for 
kinematic distance and interstellar electron density determinations 
({\it{3-5}}).  
Recently, Stanimirovic {\it{et al.}} ({\it{6}}) extended the pulsar spectral 
technique to the hydroxyl (OH) molecule with the first successful  OH
absorption measurements,  toward PSR B1849+00.  In this paper, we
expand  OH spectral measurements to eighteen additional pulsars, chosen
because they are relatively bright and lie in the inner Galaxy near the
galactic plane. One of them, PSR J1644$-$4559 = B1641$-$45, exhibits not 
only intervening OH absorption   at  1612, 1665, and 
1667 MHz, but also interstellar stimulated emission at 1720 MHz. 

The widths and strengths of spectral lines from some interstellar molecules
provide abundant indirect evidence for stimulated emission processes; indeed 
it was recognized soon after the discovery of interstellar OH in the 1960s that
maser processes must be involved ({\it{7-9}}).  Beam-switched
OH measurements toward  3C123 have also demonstrated that a background
source can stimulate emission in interstellar OH ({\it{10}}). The pulsed   
1720 MHz maser detection reported here represents  {\em{direct}} 
astronomical observation of the process of the radiative stimulation of  
emission. Here  the broadband pulsar spectrum exhibits excess line 
emission at 1720 MHz as the pulsar's photons stimulate the creation of 
additional photons in an intervening OH cloud.  This excess emission switches 
on and off with the pulsar, clearly demonstrating its stimulated nature. In 
this report, we analyze these pulsar absorption and stimulated emission  
observations and we compare them with similar measurements in non--pulsar
observations to study the physical properties of the intervening medium.

 A pulsar binning
spectrometer  was employed to ultimately create two separate spectra:  a 
``pulsar'' spectrum and a ``pulsar-{\em{off}}'' spectrum.  (See supporting 
online materials ({\it{11}}) for additional details.)  The pulsar spectrum 
represents the signal of the pulsar alone, as absorbed or amplified by 
intervening OH.  In contrast, the pulsar-{\em{off}} spectrum is sensitive to 
OH emission or absorption occurring anywhere within the telescope beam.  
The ``main'' OH lines at 1665 and 1667 MHz were simultaneously observed 
in the 4 MHz bandpass with several-hour integrations for each of the 18 
pulsars in our sample.  (See Table 1 for total integration
times and for 1--$\sigma$ optical depth uncertainties in the Hanning smoothed 
pulsar spectra.)  After our success in detecting  absorption in the 1665 and 
1667 MHz pulsar spectra of PSR B1641$-$45, we also 
measured  the ``satellite'' (1612 and 1720 MHz) lines  toward the pulsar.  

In order to calculate the optical depth $\tau$ of absorption or stimulated 
emission, consider the defining equation for optical depth: 
$I/I_o = e^{-\tau}$, where $I_o$ is the original intensity and $I$ is the 
intensity after traversal through optical depth $\tau$ 
of material.  The differencing procedure leading to the {\em{pulsar}} spectrum 
automatically eliminates all non--pulsar signals and yields the spectrum in 
units of $(I/I_o)|_{PSR}$, so it is straightforward  to  calibrate the pulsar 
spectrum in terms of  $\tau$.  It is
not so simple to determine optical depths in the pulsar-{\em{off}} spectra 
for several reasons.  First, it is not clear whether the background emission
and foreground absorption/amplification regions subtend the same solid angle.
It is probably reasonable to assume in our case that the background fills the 
telescope beam since its  predominant source is the smoothly distributed 
galactic
nonthermal emission, but the size of the foreground clouds is unknown from
our measurements.  Consequently in the absence of better information,
we will assume that both foreground and background fill the beam, so that
$I/I_o = T/T_o$.  However, another difficulty is that the observed
continuum is emitted throughout the line of sight across the Galaxy,
whereas the definition of $\tau$ requires that $T_o$  be only that portion
 emitted {\em{beyond}} the cloud contributing to optical depth. In order
to estimate the fraction of the continuum $f(d)$ contributing to $T_o$ 
as a function of distance $d$, we synthesized a  model of the continuum 
emission along the line of sight, consisting of galactic synchrotron 
({\it{12}}), ionized hydrogen regions, and the 2.7 K Cosmic  Background
Radiation.  We then integrated this model  along the 
appropriate (background) part of the line of sight, and normalized the 
result by the total emission along the line of sight:

\begin{equation}
f(d)=  \frac {\epsilon_{\rm{background}}} {\epsilon_{\rm{tot}}}
=\frac{\int_{d}^{\infty} \epsilon(s) ds }{ \int_{0}^{\infty} \epsilon(s) ds}
\ \ .
\end{equation}

Then $T_o(d)= T_b \  f(d)$, where $T_b$ is the brightness temperature, 
determined as described in supporting online materials ({\it{11}}).    
Distance was then mapped to radial velocity of the spectra with a  galactic 
rotation model ({\it{13}}).

Fig.~1 displays the 1720 MHz spectra toward PSR B1641$-$45 which directly 
demonstrate the process of stimulated emission.  The pulsar-{\em{off}} (Fig. 
$1a$, bottom)
spectrum, acquired in the interval between pulses, shows both emission and
absorption against other background source(s) lying within the 13 arcmin 
telescope beam.  The pulsar-{\em{on}} (Fig. $1a$, top)
spectrum appears to zeroth order to be merely a copy of the pulsar-{\em{off}}
spectrum, shifted upward by a constant equal to the broadband pulsar signal
strength.  However, when these two spectra are carefully differenced ({\it{11}})
 to create the pulsar
spectrum (Fig. $1b$), it is clear that there is excess signal 
at  $v \sim -45$ km/s, where the broadband pulsar signal 
has been amplified
by stimulated emission. It has long been thought that
stimulated emission plays an important role in astrophysical OH line radiation.
For example,  line temperatures  are frequently far in excess of those 
inferred from (assumed thermal) line widths ({\it{7-9}}). However, our   
measurements {\em{directly}} demonstrate the stimulated amplification 
of a signal propagating through the interstellar medium, in that the 
amplification 
is directly  observable as the pulsar cycles on and off during its 455 ms 
rotational period.  As the  stimulated emission switches on and off 
synchronously with the pulsar pulse, our 14 ms time 
resolution [ $\frac{1}{32}^{nd}$of the pulse period ({\it{14}})] places an 
amazingly short upper limit on its duration.   Since the shortest  intrinsic 
fluctuation timescale previously reported was $\sim1000$ sec ({\it{15}}),
these variations are by far the quickest observed in any interstellar maser.

Two conditions must  be satisfied in order that stimulated emission
occur: First, the relevant level populations must be inverted or ``pumped''
by some process; and second, appropriate photons should be available
to stimulate the emission from the upper, 
overpopulated level.  In our case, the
level inversion is accomplished locally in the OH cloud by a low-energy
radiative or collisional process, while the stimulating photons are
provided by the pulsar.

Note that the pulsed maser line optical depth $\tau\sim-0.05$, implying that
approximately five excess line photons are stimulated in the cloud for every
hundred passing through it. As the maser is unsaturated with a gain of only 
1.05, we expect that the FWHM of the line should be very similar to the 
expected thermal line width, which is about 0.5--0.7 km/s for gas with 
kinetic temperature of 100--200 K.  For example, 
a typical FWHM of 0.5 km/s was found in a large survey of 1720 
MHz masers in star-forming regions ({\it{16}}).
The FWHM we measure  is about 2 km/s, 
slightly wider and suggesting that we are most likely seeing a blend of 
several maser spots along the line-of-sight.

It has been suggested ({\it{17}}) that extraterrestrial civilizations
could use interstellar masers to amplify their radio transmissions.  We have
demonstrated here that such a process could sustain modulation down to 
millisecond
timescales, but of course the  gain of this particular maser is too small
to provide significant amplification of an ETI signal.

PSR B1641$-$45 = J1644$-$4559 lies in a well--studied ({\it{18-22}}) region 
of the inner Galaxy near the galactic plane,
at galactic longitude $l$ and latitude $b = (339.2\dg, -0.2\dg)$. We are able
to construct a  schematic map of the interstellar medium along the line of 
sight by combining our observations with earlier ones (see Fig. 2).

On the basis of kinematic analysis of  HI absorption 
spectra  ({\it{23}}), the pulsar is 
placed 4.6 kpc along  the line of sight. Two  ionized 
hydrogen (HII) regions lie in this direction, with galactic
coordinates and recombination line velocities  ({\it{19,20}}) of
$(l,b,v_{\rm{LSR}})=(339.1\dg,
-0.2\dg, -120$ km/s) and $(339.1\dg, -0.4\dg, -37$ km/s). With  the
rotation curve  of ({\it{13}}), our kinematic 
analysis of the recombination line measurements  places
G339.1--0.2 beyond the pulsar at a geocentric distance $d\sim 6.7$ kpc, and 
G339.1--0.4 closer than the pulsar at $d\sim3.3$ kpc.

Fig. 3 displays  spectra at frequencies of the four 
ground rotational state 18 cm OH lines toward PSR B1641$-$45.
The most interesting feature 
is a strong spectral line at $v_{\rm{LSR}}\sim-45$ km/s in all our 
OH spectra -- both pulsar-{\em{off}} and pulsar spectra.  
The line is in absorption at 1612, 1667, 
and 1665 MHz and in  emission at 1720 MHz (the latter being the pulsed maser
emission discussed above).   As it is visible in
the pulsar spectra (right column, Fig. 3), this line must arise {\em{between}}
the pulsar and the observer. It probably originates in OH gas associated 
with or near G339.1--0.4, as the velocities are similar. Note however that 
{\it{pervasive}} extended regions of 1720 MHz emission have been found in 
the inner galactic plane ({\it{24,25}}), including a  $>1\dg$--long filament 
crossing near the pulsar line of sight with $v_{\rm{LSR}}\sim-40$ km/s.  
As our 1720 MHz pulsar-{\em{off}} spectral line at $-45$ km/s has  
strength and width similar to the 
extended OH gas (though the velocities are somewhat discrepant),  it is 
possible that it originates from this extended 
OH region rather than from the HII region G339.1--0.4.

The --45 km/s emission and absorption features show  evidence for departures
from local thermodynamic equilibrium (LTE). The pulsar-{\em{off}} spectra 
typically  have FWHM of 2-3 km/s, which is a few times wider than what is 
expected for thermally broadened line profiles at a typical kinetic 
temperature of about 100 K.  The peak optical depths  are very similar at 
1667 and 1665 MHz ($\tau\sim$0.03),  while in the LTE case their ratio would 
be $9/5$.  In addition, the 1612 MHz lines are inverted with respect to  1720 
MHz.

We  see another OH line  at  $v_{\rm{LSR}}\sim-30$ km/s in most of the 
eight spectra toward PSR B1641$-$45 shown in Fig. 3,
including the 1665 and 1667 MHz pulsar
spectra. This line must therefore also originate in gas  nearer than
the pulsar  to us, again probably associated with or near G339.1--0.4.
The lines are seen in absorption in  pulsar spectra and primarily in emission
in  pulsar-{\em{off}} spectra.  

Finally, all pulsar-{\em{off}} spectra 
exhibit line(s) at $v_{\rm{LSR}}\sim -100$ to --120 km/s, which are not seen
in the pulsar spectra.  Consequently  they must originate in gas beyond the 
pulsar, probably associated with or near G339.1--0.2.

Note that each 1720 MHz spectrum (Fig. 3, top row)  is an inverted copy of the 
1612 MHz spectrum (Fig. 3, second row). This phenomenon,  called
``conjugate'' line behavior, occurs because the initial states of both 
transitions are overpopulated by an identical 
process ({\it{8,26-28}}). For our predominantly observed conjugate state, 
having 1720 MHz stimulated emission and  1612 MHz stimulated absorption,  the 
process begins in a region having $T \sim 100$ K and $n_{OH} \sim 10^5$ cm$^{-3}$
with the collisional excitation of the molecule to a higher 
rotational state at energy $E = 1.66\times10^{-14}$ erg
above ground level, after which it can radiatively decay {\em{with equal 
probability}} (if the transition is optically thick) to overpopulate  either 
the upper level of the 1720 MHz 
transition or the lower level of the 1612 MHz transition.  The rarer (in our 
data) 1612 MHz
stimulated emission and conjugate 1720 MHz stimulated absorption results from a
similar process that overpopulates the opposite 18 cm levels via an 
intermediate excited rotational level  at  $E = 2.5 \times10^{-14}$ erg
above the ground level.  

Our predominant conjugate configuration becomes optically thick to 
far-infrared photons for OH column densities per velocity interval
$(N_{OH}/\Delta v) > (10^{14}$ cm$^{-2}$ s / km) , whereas the inverse
configuration becomes optically thick and then dominates  at 
$(N_{OH}/\Delta v) > 
(10^{15}$ cm$^{-2}$ s / km). Hence our predominant conjugate
configuration (including the 1720 MHz  pulsar-stimulated emission and 
the  pulsar-{\em{off}} emission at $v\sim-120$ km/s)
originates in clouds with specific column densities between these two 
limits, while the rarer opposite configuration ({\it{e.g}}., pulsar-{\em{off}}
1720 MHz absorption at $v\sim-100$ km/s) originates in a column whose 
density is above the upper limit.  Then the occasionally observed 
{\it{adjacent}}  emission and absorption features  that
are conjugate at the two frequencies ({\it{e.g}}., the pulsar-{\em{off}}
spectra at $v\sim-32$ km/s) suggest a   density gradient in the cloud 
({\it{28}}), with specific column densities crossing  
$(10^{15}$ cm$^{-2}$ s / km) at the transition.

It is useful to compare the lines observed in the pulsar and pulsar-{\em{off}} 
spectra, since all were acquired at the same time with the telescope pointing
in exactly the same direction. To facilitate the comparison, optical depth
scales on the right side of all eight specra in Fig. 3 are identical.
 The $v\sim-45$ km/s lines  exhibit markedly 
stronger $(|\tau|\sim(2-3) \times$ larger) absorption and stimulated emission 
in  pulsar spectra (Fig. 3, right column) than in 
the corresponding pulsar-{\em{off}} spectra  (Fig. 3, left column).
The discrepancy at $v\sim-32$ km/s is even stronger --
absorption in pulsar spectra at 1665 and 1667 MHz is absent or replaced by 
weak {\em{emission}} in pulsar-{\em{off}} spectra  ({\it{29}}).  These results 
are striking, especially since the only 
other successful pulsar OH absorption experiment  ({\it{6}}) also found 
stronger absorption in the pulsar spectra than in the pulsar-{\em{off}} 
spectra.   
 
The {\em{widths}} of the lines  are narrower in our pulsar spectra than in 
pulsar-{\em{off}} spectra at 1720 and 1612 MHz, but similar at 1667 and
1665 MHz.  The earlier results ({\it{6}}) exhibited narrower lines  
in the pulsar spectra than in the pulsar-{\em{off}}  spectra at 1667 and
1665 MHz.

Our new observations strengthen 
the earlier interpretation that the needle--thin interstellar column sampled 
by the pulsar signal interacts with a 
substantially different sample of the medium than does the pulsar-{\em{off}} 
column, which
represents the average of all interactions across the 13 arcmin 
telescope beam. 
Presumably the pulsar signal is encountering  small, dense OH cloudlets 
whose properties
are diluted in the beam--averaged pulsar-{\em{off}} spectrum.  This behavior 
differs markedly from HI, where the statistics  show {\em{no}} dependence on 
the  angular cross--sections of absorbing columns across a
tremendous range of solid angle ({\it{30,31}}).  Interestingly, molecular gas
is known to be more clumped than neutral gas, at least on larger scales.

If  the difference in OH optical depths indeed results from clumping, we would
expect that other pulsar lines of sight would pierce regions devoid of
cloudlets and show  {\em{shallower}} optical depths than 
pulsar-{\em{off}} measurements. One might ask why  there are no such
complementary results.  Unfortunately an observational selection effect
conspires against success in such observations, as we do not have sufficient
sensitivity on any other pulsar lines of sight (see Table 1) to test this 
hypothesis.

An earlier paper  ({\it{32}}) reported broad $(\Delta v> 10$ km/s) and deep 
$(\tau > 0.5)$ OH absorption at 1667 MHz in the spectrum of PSR B1749$-$28. We 
have adequate sensitivity to detect such a line (see Table 1), but we do not
confirm the result.  All other OH lines detected in pulsar spectra are
much narrower than the previously claimed detection in PSR B1749$-$28.

We have searched for OH absorption and stimulated emission in the 
spectrum of 18 pulsars. One pulsar, B1641$-$45, exhibits absorption or 
stimulated emission in pulsar spectra at each of the four 18 cm line 
frequencies. No absorption or 
stimulated emission was detected in the others, including 
one in which OH  absorption had previously been reported (B1749$-$28).
 A variety of interesting results are  drawn from the B1641$-$45 
spectra. The pulsed maser line, with $\tau\sim-0.05$, represents
the first direct  detection of interstellar stimulated emission.
The  OH and HII concentrations are mapped along the line of sight to the 
pulsar, and they are found to be associated kinematically and probably 
spatially.  Analysis of the lines provides insight into the OH density, 
temperature, and excitation. Finally, the relative depths of lines in pulsar
spectra and pulsar-{\em{off}} spectra suggest that the OH gas is highly 
clumped. 

\section*{Supporting online material: Methods}

The  observations reported here were obtained with the 64-m telescope 
located near Parkes, NSW, Australia, in September 2004, with the  H-OH 
front--end receiver package.  At our 18-cm observing wavelength, 
the telescope has a sensitivity of 1.5 Jy/K 
and a beam diameter of 13 arcmin. The
back--end correlation spectrometer was used in the pulsar binning mode, 
in which each correlation function was integrated into one of thirty--two 
pulse phase bins. Each of the 32  resulting spectra covered a 4 MHz band that
was divided into 2048 spectral channels, each $\sim$2 kHz or $\sim$0.4 km/s 
wide. The spectra acquired in
the phase bins corresponding to the pulsar pulse were collapsed into a single
pulsar-{\em{on}} spectrum; while the spectra gathered in the interval between 
pulses were integrated into a single pulsar-{\em{off}} spectrum. The 
{\em{pulsar}} spectrum was formed from differencing pulsar-{\em{on}} and 
pulsar-{\em{off}} spectra, and normalizing
by the mean pulsar temperature.  No additional baseline flattening is 
required in the pulsar spectrum, while low-order sinusoids were fitted to and
removed from  the final pulsar-{\em{off}} spectrum.  
Lastly,  all spectra were Hanning smoothed, giving a final resolution of 
$\sim$0.7 km/s. We have previously used similar pulsar binning 
spectrometric techniques for HI measurements at Parkes and Arecibo
observatories ({\it{1-5}}), and for OH at Arecibo ({\it{6}}).

The data were calibrated using observations of Hydra~A, whose flux density
near 1660~MHz was assumed to be 36.8~Jy.   The system temperature  
$T_{sys}$ was found to be 28.7 and 29.9 K  in the two linearly 
polarized feed probes.  An additional contribution from the sky at the pulsar
position  was estimated by convolving a 1400-MHz interferometric image 
({\it{22}}) with a 13 arcmin beam and then extrapolating the
resultant brightness temperature (less contributions from the Cosmic
Background Radiation  and
HII regions) to 1650~MHz with a spectral index
 of -2.7. Line brightness temperatures were then determined under the 
 assumption that the line--emitting regions fill the telescope beam,
 by measurement of the line--to--total continuum ratio.

\begin{quote}
{\bf References and Notes}


\item  1.   J.~M. Cordes,  J.~M. Weisberg, V. Boriakoff, {\it{Astrophys. J.}} 
{\bf{288}}, 221 (1985).

\item  2. S. Johnston, L. Nicastro, B. Koribalski, 
{\it{Mon. Not. R. Astron. Soc.}} {\bf{297}}, 108 (1998).

\item  3.  B. Koribalski, S. Johnston,  J.~M.  Weisberg, W. Wilson, 
 {\it{Astrophys. J.}} {\bf{441}}, 756 (1995).

\item 4.  J.~M.  Weisberg,  M.~H. Siegel,  D.~A. Frail, S. Johnston,  
{\it{Astrophys. J.}} {\bf{447}}, 204  (1995).
 
\item  5.  S.  Johnston, B. Koribalski, B.,  J.~M. Weisberg, W.  Wilson,   
{\it{Mon. Not. R. Astron. Soc.}} {\bf{322}}, 715 (2001).

\item  6.  S. Stanimirovi{\' c},  J.~M.  Weisberg, J.~M. Dickey, A. de la 
Fuente, K. Devine, A. Hedden,  S.~B. Anderson, {\it{Astrophys. J.}} {\bf{592}},
953 (2003).

\item 7. A. H. Cook,{\it{ Celestial Masers}} (Cambridge Univ. Press, Cambridge,
1977).

\item  8. M. Elitzur, {\it{Astronomical Masers}} (Kluwer, Dordrecht, 1992).

\item 9. M.~ J.  Claussen, {\it{Science}} {\bf{306}}, 235 (2004).

\item 10. N.~Q. Rieu, A.  Winnberg,  J.  Guibert,  J.~R.~D.  Lepine, L.~E.~B.  
Johansson, W.~M.  Goss,  {\it{Astron. Astrophys.}}  {\bf{46}}, 413 (1976).

\item  11.  The observing technique is supplied as supporting material on 
{\it{Science}} Online. It is also the last section of this preprint before references.

\item   12. K. Beuermann, G. Kanbach, E. M. Berkhuijsen,   {\it{Astron.
 Astrophys.}}  {\bf{153}}, 17 (1985).

\item    13. M. Fich, L. Blitz, A. A. Stark,  {\it{Astrophys. J.}} {\bf{342}}, 
272, (1989). 

\item  14.  The pulsar binning spectrometer was adjusted so that the pulsar 
pulse fell entirely within one of the 32 phase bins ({\it{11}}). Consequently 
no information on possible shorter timescale phenomena is available.

\item  15.  A.~W. Clegg, J.~M. Cordes,  {\it{Astrophys. J.}} {\bf{374}}, 150 
(1991).

\item  16.  J.~L. Caswell,  {\it{Mon. Not. R. Astron. Soc.}} {\bf{349}}, 99 
(2004).

\item  17.  J. Cordes,  {\it{Third Decennial US-USSR Conference on SETI,}} 
Ed. G.~S. Shostak, {\it{Astron. Soc. Pacific Conf. Series}} {\bf{47}}, 257 
(1993).

\item   18. R.~N. Manchester, U. Mebold,  {\it{Astron. Astrophys.}}  {\bf{59}}, 
401 (1977).

\item   19. P.~A. Shaver, R.~X. McGee, L.~M. Newton, A.~C. Danks, S.~R. 
Pottasch, {\it{Mon. Not. R. Astron. Soc.}} {\bf{204}}, 53 (1983).

\item   20. J.~L. Caswell,  R.~F. Haynes,   {\it{Astron. Astrophys.}}  
{\bf{171}}, 261 (1987).

\item   21. J.~C. Cersosimo, {\it{Astrophys. J.}} {\bf{349}}, 67 (1990).

\item   22. N.~M. McClure-Griffiths,  J.~M. Dickey, B.~M. Gaensler,  A.~J. 
Green,  M. Haverkorn, S. Strasser,  {\it{astro-ph/}} 0503134 (2005)

\item   23. D.~A. Frail, J.~M. Weisberg,   {\it{Astron. J.}}  {\bf{100}}, 
743 (1990).

\item  24.  R.~F. Haynes, J.~L. Caswell, {\it{Mon. Not. R. Astron. Soc.}} 
{\bf{178}}, 219 (1977).

\item   25. B.~E. Turner,  {\it{Astrophys. J.}} {\bf{255}}, L33 (1982).

\item   26. H.~J. van Langevelde,  E.~F. van Dishoeck,  M.~N. Sevenster,  F.~P.
Israel, {\it{Astrophys. J.}} {\bf{448}}, L123 (1995).

\item  27.  P. Lockett, E. Gauthier, M. Elitzur,  {\it{Astrophys. J.}} 
{\bf{511}}, 235 (1999).

\item   28. K.~J. Brooks,  J.~B. Whiteoak, {\it{Mon. Not. R. Astron. Soc.}}
{\bf{320}}, 465 (2001).

\item  29. Note that the $v\sim-100$ km/s lines in the  pulsar-{\em{off}} 
spectra can not have  analogs in the pulsar spectra since they originate in 
gas beyond the pulsar.

\item  30.  J.~M. Dickey, J.~M. Weisberg, J.~M. Rankin, V. Boriakoff,  
{\it{Astron. Astrophys.}}   {\bf{101}}, 332 (1981).

\item  31.  H.~E. Payne, Y. Terzian,  E.~E. Salpeter, {\it{Astrophys. J. 
Suppl. Ser.}} {\bf{48}}, 199 (1982).

\item  32. V.~I. Slysh, {\it{Astronomicheskii  Cirkular}}, {\bf{731}}, 1 (1972).

\item   33. The quantity  $\sigma_{\tau}$ (pulsar spectrum) in Table 1 is
the optical depth standard deviation in the Hanning smoothed pulsar spectrum, 
which includes the effects of radiometer and sky noise, and in some cases, 
interstellar scintillation.

\item We thank  John Reynolds 
and Warwick Wilson for assistance with the gated correlator configuration, 
K. Wells and K. Willett  for help  with the observations, and  R. Norris 
and J. Caswell for providing  useful suggestions.   JMW  gratefully 
acknowledges financial support from NSF grant AST 0406832, the Australia 
Telescope National Facility, and the School of Physics of
the University of Sydney. SS acknowledges support 
by NSF grants  AST 0097417 and AST 9981308. The  Parkes telescope
is part of the Australia Telescope which is funded by the Commonwealth of 
Australia for operation as a National Facility managed by CSIRO.

\end{quote}

{\bf{Table 1:}} Integration times and pulsar spectrum noise fluctuations.\\
\\
\begin{tabular}{|l|l|c|c|l|}
\hline
PSR J & PSR B & Freq  & $t_{tot}$ & $\sigma_{\tau}$  ({\it{33}}) \\
      &       & (MHz) &    (hr)   &  (pulsar                      \\
      &       &       &           &   spectrum)                  \\
\hline
\hline
0742--2822 & 0740--28 & 1665/7 & 3 & 0.1 \\
0835--4510 & 0833--45 & 1665/7 & 2 & 0.1 \\
0837--4135 & 0835--41 & 1665/7 & 4 & 0.07\\
0908--4913 & 0906--49 &  1665/7 &4 & 0.1 \\
1056--6258 & 1054--62 &  1665/7 &2 & 0.1 \\
1057--5226 & 1055--52 &  1665/7 &1 & 0.1 \\
1157--6224 & 1154--62 &  1665/7 &2 & 0.3 \\
1243--6423 & 1240--64 &  1665/7 &2 & 0.1 \\
1326--5859 & 1323--58 &  1665/7 &2 & 0.1 \\
1327--6222 & 1323--62 &  1665/7 &2 & 0.1 \\
1600--5044 & 1557--50 &  1665/7 &7 & 0.1 \\
1605--5257 & 1601--52 &  1665/7 &1 & 0.2 \\
1644--4559 & 1641--45 &  1665/7 &5 & 0.01 \\
 \ \ ``  \ \ \ \ \ \ \   ``     & \  \ ``  \ \ \ \ \ \ \   ``    &  1720   &5 & 0.01 \\
 \ \ ``  \ \ \ \ \ \ \   ``     & \  \ ``  \ \ \ \ \ \ \   ``    &  1612   &4 & 0.01 \\
1745--3040 & 1742--30 &  1665/7 &5 & 0.2 \\
1752--2806 & 1749--28 &  1665/7 &4 & 0.05\\
1803--2137 & 1800--21 &  1665/7 &5 & 0.3 \\
1825--0935 & 1822--09 &  1665/7 &1 & 0.3 \\
1829--1751 & 1826--17 &  1665/7 &2 & 0.3 \\

\hline

\end{tabular}

\newpage

\begin{figure}[ht]
\centering
\resizebox{14 cm}{!}{\includegraphics{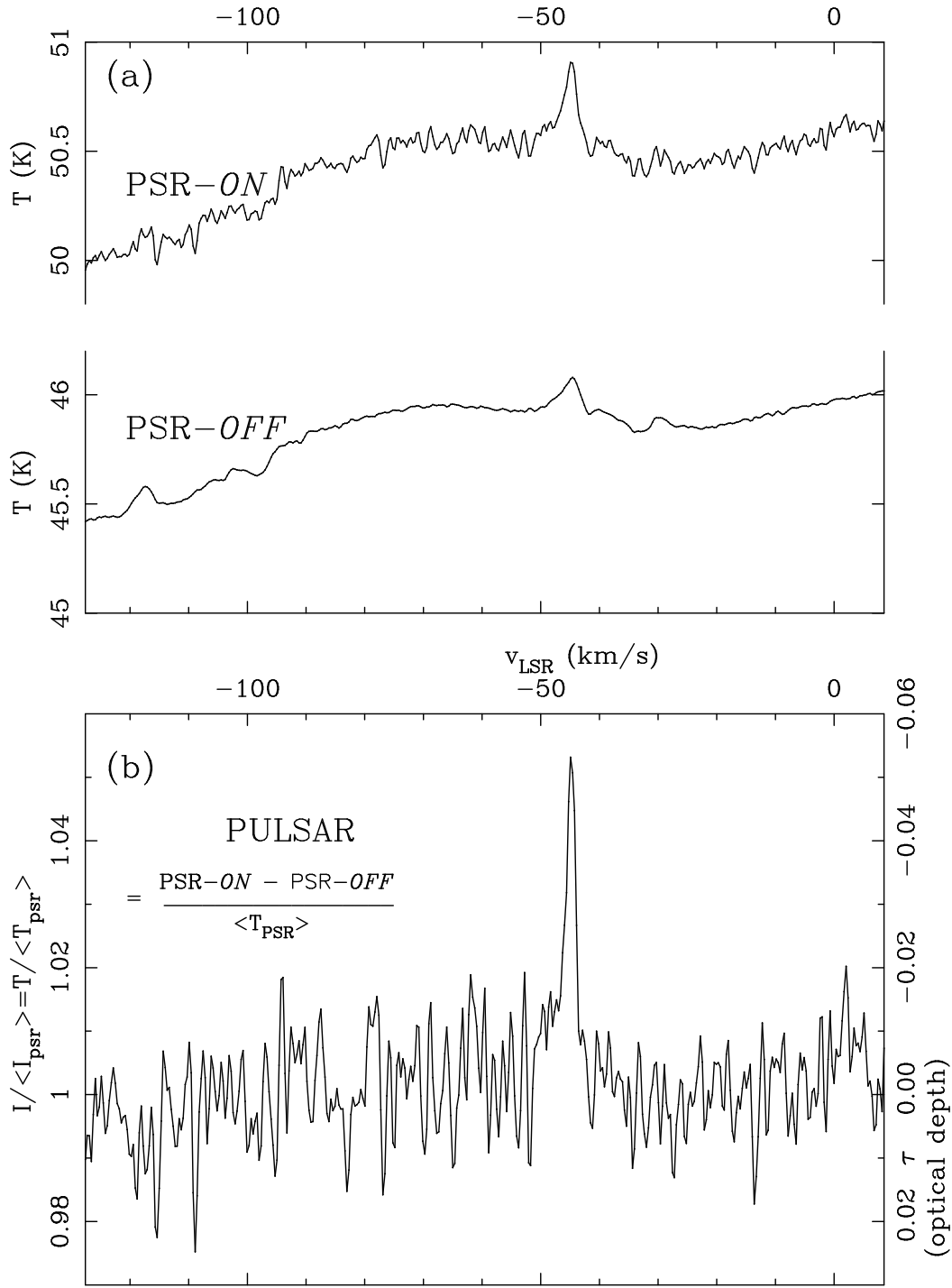}  }
\caption{Stimulated amplification of the PSR B1641$-$45  
signal   in an interstellar OH cloud at 1720 MHz. 
(a). The ``pulsar-{\em{on}}''spectrum, acquired during 
the pulsar pulse, and the 
``pulsar-{\em{off}}'' spectrum, gathered in the interval 
between pulses. The two spectra exhibit both emission and
absorption against other (non--pulsar) background source(s) lying 
within the 13 arcmin telescope beam, while the pulsar-{\em{on}} 
spectrum additionally contains the pulsar signal. 
(b). The ``pulsar'' spectrum, the difference of pulsar-{\em{on}} 
and pulsar-{\em{off}},
illustrating the pulsar signal {\em{alone}} as absorbed (or in this 
case, amplified) by 
intervening OH. The spike in this spectrum at $v_{\rm{LSR}}\sim-45$ 
km/s results 
from excess emission in an OH cloud, stimulated by pulsar photons.   }
\end{figure}

\begin{figure}[ht]
\centering
\resizebox{15 cm}{!}{\includegraphics{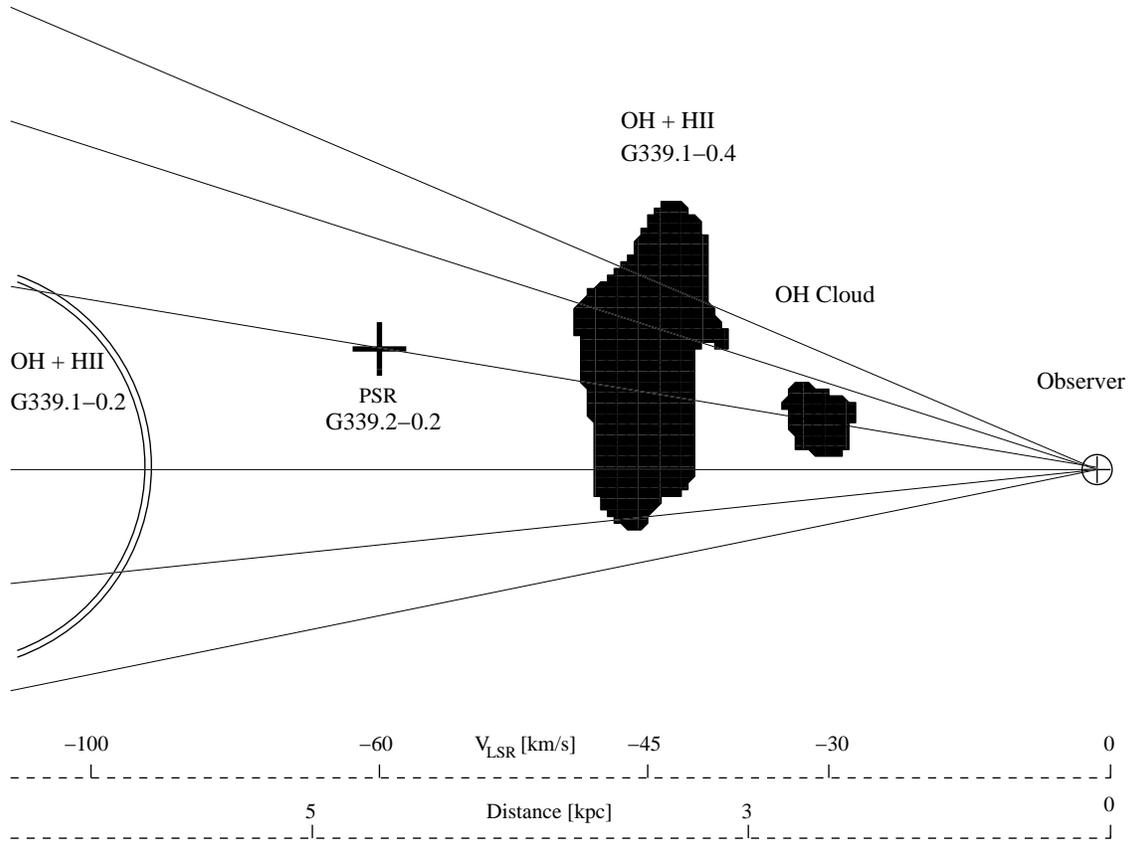}  }
\caption{  A schematic model of the ISM toward PSR B1641$-$45. The 
ionized region 
velocities are  from ({\it{19,20}}), the OH velocities are from
the current work,  and the limiting pulsar HI 
absorption velocities
are from ({\it{23}}). The kinematic velocity--to--distance 
conversion  uses the rotation curve of ({\it{13}}). The lines represent 
various lines of sight within the 13 arcmin telescope beam. The vertical 
scale has been enlarged for clarity.}
\end{figure}

\begin{figure}[ht]
\centering

\resizebox{14 cm}{!}{\includegraphics{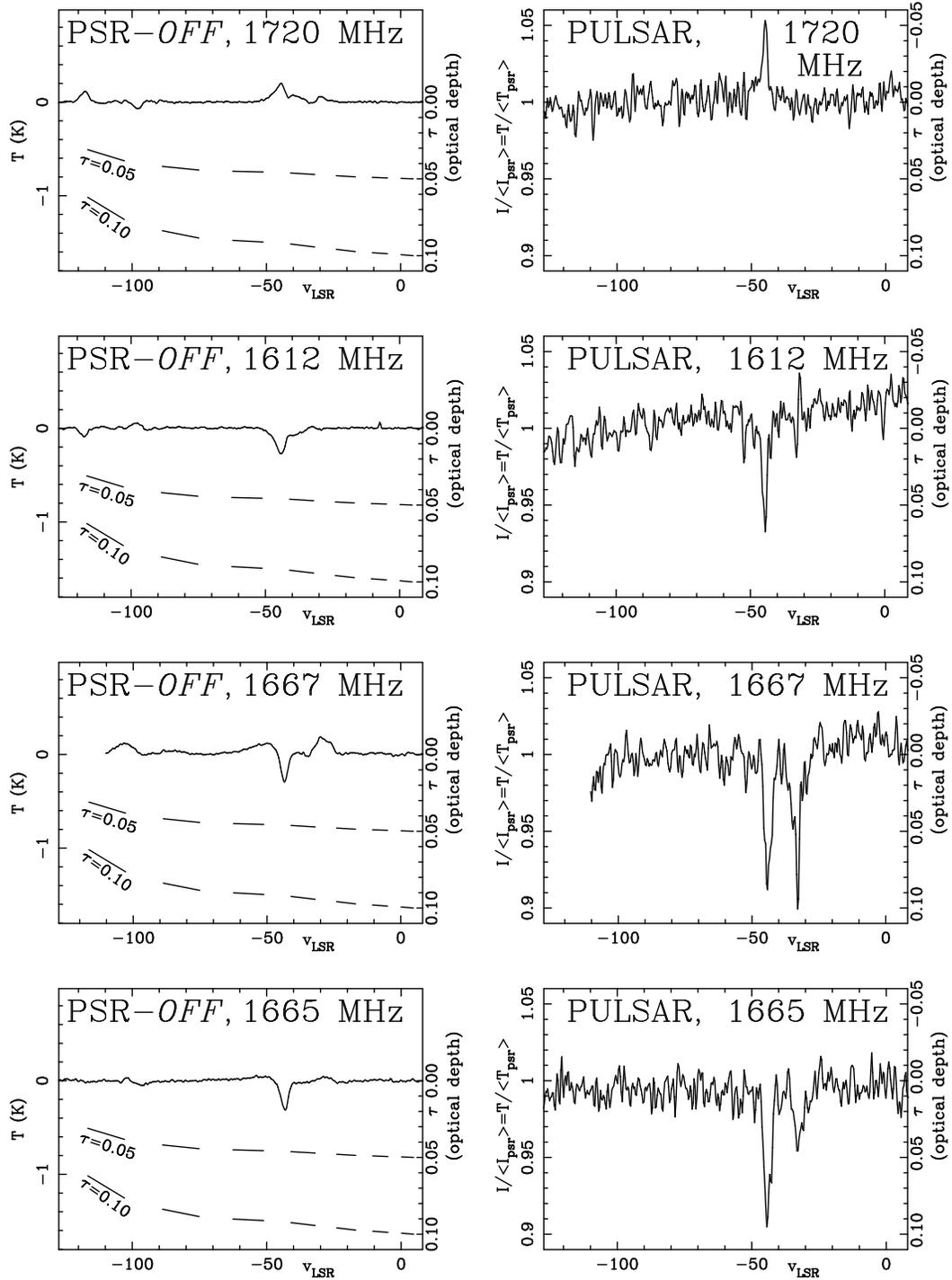}  }

\caption{Spectra of the four 18 cm ground--state rotational 
transitions of OH toward PSR B1641-45.  The left column displays 
the four ''pulsar-{\em{off}}'' spectra, which are sensitive to 
all emission and absorption in the 13 arcmin 
telescope beam when the pulsar is switched off.  The right 
column shows the four ``pulsar'' spectra, which exhibit the 
absorption or stimulated emission of the  pulsar signal alone.  
The right--hand ordinate on each panel is optical depth $\tau$; 
all eight spectra are plotted with the same 
optical depth scale.  Note that all pulsar-{\em{off}} spectral 
features are significantly shallower (i.e., have smaller optical
depths) than their analogs in the pulsar spectra.  In the pulsar-{\em{off}} 
spectra, the sloping lines of constant optical depth result from changes  
in the  ratio of background to total continuum along the line of sight
 (see discussion accompanying Eq. 1). Low--order 
sinusoids were fitted to and removed from 
the pulsar-{\em{off}} baselines in order to flatten them. }

\end{figure}

\end{document}